\documentclass[journal,twoside,web]{ieeecolor}
\usepackage{generic}
\usepackage{amsfonts,mathtools}
\usepackage{amssymb}
\usepackage{amsmath}
\usepackage{graphicx}
\usepackage[all]{xy}

\usepackage{color}
\usepackage{xcolor}
\usepackage{bm}
\usepackage{csquotes}
\usepackage[section]{algorithm}

\DeclareMathOperator{\lead}{\text{lead}}
\DeclareMathOperator{\init}{in}
\DeclareMathOperator{\rank}{rank}
\DeclareMathOperator{\tr}{tr}

\renewcommand{\thesection}{\Roman{section}}
\renewcommand{\thesubsection}{\Alph{subsection}}

\usepackage[colorlinks, linkcolor=reference,
 citecolor=citation,
         urlcolor=e-mail]{hyperref}

\RequirePackage{verbatim}

\RequirePackage{hyperref}

\RequirePackage{color}

\definecolor{conditional}{rgb}{0,1,0}
\definecolor{e-mail}{rgb}{0,.40,.80}
\definecolor{reference}{rgb}{.20,.60,.22}
\definecolor{mrnumber}{rgb}{.80,.40,0}
\definecolor{citation}{rgb}{.20,.60,.22}

\usepackage{bbm}

\usepackage{tikz}
\usetikzlibrary{arrows,decorations.markings}

\newtheorem{theorem}{Theorem}

\newtheorem{corollary}{Corollary}

\newtheorem{definition}{Definition}
\newtheorem{remark}{Remark}
\newtheorem{example}{Example}

\begin{document}
\title{Algorithm to find new identifiable reparametrizations of parametric rational ODE models}

\author{Nicolette Meshkat, Alexey Ovchinnikov, and Thomas Scanlon
\thanks{This work was partially supported by the NSF grants CCF-2212460, CCF-1563942, CCF-1564132, DMS-1760448, DMS-1760212, DMS-1760413, DMS-1853650, and DMS-1853482 and CUNY grant PSC-CUNY \#65605-00 53.}
\thanks{Nicolette Meshkat is with Santa Clara University, Department of Mathematics and Computer Science, 500 El Camino Real,
Santa Clara, CA 95053, USA (e-mail: nmeshkat@scu.edu).}
\thanks{Alexey Ovchinnikov is with CUNY Queens College, Department of Mathematics,
65-30 Kissena Blvd, Queens, NY 11367, USA and 
CUNY Graduate Center, Mathematics and Computer Science, 365 Fifth Avenue,
New York, NY 10016, USA (e-mail: aovchinnikov@qc.cuny.edu).}
\thanks{Thomas Scanlon is with University of California, Berkeley, Mathematics Department, Evans Hall, Berkeley, CA, 94720-3840 (e-mail: scanlon@math.berkeley.edu).}}

\maketitle

\begin{abstract}
Structural identifiability concerns the question of which unknown parameters of a model can be recovered from (perfect) input-output data.  If all of the parameters of a model can be recovered from data, the model is said to be identifiable.  However, in many models, there are parameters that can take on an infinite number of values but yield the same input-output data.  In this case, those parameters and the model are called unidentifiable.  The question is then what to do with an unidentifiable model.  One can 
try to add more input-output data or decrease the number of unknown parameters, if experimentally feasible, or try to find a reparametrization to make the model identifiable.  In this paper, we take the latter approach. While existing approaches to find identifiable reparametrizations were limited to scaling reparametrizations or were not guaranteed to find a  globally identifiable reparametrization even if it exists, we 
 significantly broaden the class of models for which we can find a globally 
identifiable model with the same input-output behavior as the original one.  We also prove that, for linear models, a globally identifiable reparametrization always exists and show that, for a certain class of linear compartmental models, with and without inputs, an explicit reparametrization formula exists.  We illustrate our method on several examples and provide detailed analysis in supplementary material on github.
\end{abstract}


\section{Introduction}
Structural (local) identifiability is a property of an ODE model with parameters
\begin{equation}\label{eq:mainsys}
\begin{cases}
\bar x'(t) = \bar f(\bar x(t),\bar\alpha,\bar u(t))\\
\bar y(t) = \bar g(\bar x(t), \bar\alpha, \bar u(t)),
\end{cases}
\end{equation}
 as to whether the parameters $\bar\alpha$ can be uniquely  determined (or determined up to finitely many choices) from the inputs $\bar u$ and outputs $\bar y$ of the model. If a parameter is not locally identifiable, then it is not possible to estimate its numerical values from measurements of the outputs. Non-identifiability occurs rather frequently in models used in practice~\cite{BV2023}. Therefore, it is important to develop theory and algorithms that can eliminate non-identifiability.  Achieving  only local identifiability for a model (finitely many parameter values fit the data) can still be problematic for many algorithms and software packages for parameter estimation. This is because these algorithms typically cannot find all of the multiple parameter values that fit the data, and multiple values can fit into the physically meaningful ranges~\cite{BV2023}. As a result, errors in such methods can easily be much higher than for globally identifiable models, see the locally identifiable Biohydrogenation, Mammillary~4, and SEIR models in the tables in~\cite{par_est_robust}. Therefore, it is important to find a globally rather than just locally identifiable reparametrization.
 
 In this paper, we discuss closely related properties called  global and local input-output (IO) identifiability, which concern determining the parameters from IO-equations,  i.e.  the equations relating the inputs and the outputs obtained by eliminating the state variables~\cite{O1990,O1991,DJNP2001,MRCW21}. Global (resp., local) IO-identifiability and global (resp., local) identifiability are not logically equivalent. However, there are sufficient conditions for the equivalence, see \cite{ident-compare}, which can be checked algorithmically and often (but not always) hold in practical models.

 We propose a new method of reparametrizing an ODE model to achieve at least local structural IO-identifiability of the parameters of the reparametrized system. Whenever possible within the framework of our approach, this allows us to find a globally IO-identifiable reparametrization. However, there are ODE models for which no globally IO-identifiable reparametrizations exist regardless of the approach taken, see \cite[Section~IV.A]{OPPS2023}. {\sc Maple} code for our illustrating examples can be found in~\cite{exaples-github}. We also prove  a new general result  that, for linear models with or without inputs, a globally IO-identifiable reparametrization always exists. Additionally, for a class of linear compartmental models  with and without inputs, we obtain explicit reparametrization formulas.

 Identifiability of the initial conditions of a model, which is related to observability of the state variables, is also important. There is a new approach being developed and tested via simplifying Lie derivatives to find globally observable reparametrizations~\cite{DPR2024}. Reparametrized models in this approach are differential-algebraic systems, i.e. are of the form~\eqref{eq:mainsys} with possible additional polynomial constraints. Our algorithm cannot tackle state observability because  state variables do not appear in the IO-equations. However, reparametrized models in our approach are of the ODE form~\eqref{eq:mainsys}.

Efficient algorithms are available for finding scaling~\cite{Hubert2013} or, more generally, linear reparametrizations~\cite{Ovchinnikov2021,JimnezPastor2022}.
Further refinements are available for scaling reparametrizations of linear compartmental models~\cite{MESHKAT201446,Baaijens2015OnTE}.
Several approaches have been proposed for producing locally identifiable reparametrizations~\cite{Gunn1997,https://doi.org/10.1002/rnc.5887,JOUBERT2020108328}, which succeed in finding nontrivial parametrizations for models from the literature but are not guaranteed to produce a reparametrization if it exists.
Another recent approach~\cite{OPPS2023} gives an algorithm for reparametrizing the model preserving its structure: the reparametrized system has the same number of equations and state variables as the
old system and
the monomials in the new system are obtained from the monomials of the old system by
replacing the old state variables with the new state variables. 
This approach has shown to be practical in many cases. 
However, it has a noticeable drawback. In particular, the requirement in~\cite{OPPS2023} to  preserve the structure can result in not being able to find a globally IO-identifiable reparametrization when it exists; see \cite[Section~IV.B]{OPPS2023}   or Section~\ref{sec:bioh} from our paper for  examples, which is a limitation of that approach that we do not have in our proposed approach.

The paper is organized as follows. In Section~\ref{sec:prob2}, we state the reparametrization problem precisely. Basic definitions, including IO-identifiability, are given in Section~\ref{sec:defs}. Our main algorithm is in Section~\ref{sec:2}. We illustrate the algorithm in Section~\ref{sec:examples} using toy models, a Lotka-Volterra model with input, a chemical reaction network model, a biohydrogenation model, which is rational (non-polynomial), a bilinear model with input, and a linear compartmental model for which no scaling reparametrization exists. In Section~\ref{sec:linear}, based on our algorithm, we establish the existence of  globally IO-identifiable reparametrizations for linear models, and we also provide new general explicit reparametrization formulas, which we discovered using our software.

\section{Problem statement}\label{sec:prob2}
Given an ODE system
\begin{equation}\label{eq:ODEsystem2}
\Sigma(\bar\alpha) :=\begin{cases}
\bar x' = \bar f(\bar x,\bar\alpha,\bar u)\\
\bar y = \bar g(\bar x, \bar\alpha, \bar u),
\end{cases}
\end{equation}
where $\bar f$ and $\bar g$ are rational functions over $\mathbb{Q}(\bar\alpha)$,
find 
$\bar\beta$ in the algebraic closure of $\mathbb{Q}(\bar{\alpha})$ and $\bar{w}$ in the algebraic closure of $\mathbb{Q}(\bar x, \bar\alpha, \bar u, \bar u', \bar u'',\ldots)$ such that
\begin{itemize}
\item there exist $\bar{F}, \bar{G}$ in $\mathbb{Q}(\bar w, \bar\beta, \bar u, \bar u', \bar u'',\ldots)$ with $|\bar{F}| = |\bar{w}|$ and $|\bar{G}| = |\bar{y}|$ (we write $|\bar F|$ for
the length of the tuple $\bar F$) such that
\begin{equation}\label{eq:ODEsystem2-repar}
\begin{cases}
\bar{w}' = \bar F(\bar w,\bar\beta,\bar u, \bar u', \bar u'',\ldots)\\
\bar y = \bar G(\bar x, \bar\beta,\bar u, \bar u', \bar u'',\ldots).
\end{cases}
\end{equation}
We will denote this system by $\tilde\Sigma(\bar\beta)$.
\item all parameters $\bar\beta$ in $\tilde\Sigma(\bar\beta)$
are at least locally IO-identifiable and
\item the IO-equations of $\Sigma(\bar\alpha)$ and $\tilde\Sigma(\bar\beta)$ are the same.
\end{itemize}

 Sometimes in the literature, the ground field is taken 
to be $\mathbb{C}$, $\mathbb{R}$, or $\overline{\mathbb{Q}}$ instead of $\mathbb{Q}$.  The reader may substitute 
 these larger fields
for $\mathbb{Q}$ everywhere in this paper.  We prefer to work 
with the rational numbers as they are more amenable to machine computations.

\section{Definitions and notation}\label{sec:defs}
In this section, we recall the standard terminology from differential algebra that is used in working with IO-identifiability.
\begin{enumerate}
  \item A {\em differential ring} $(R, {}')$ is a commutative ring with a derivation $':R\to R$, that is, a map such that, for all $a,b\in R$, $(a+b)' = a' + b'$ and $(ab)' = a' b + a b'$. 
  \item The {\em ring of differential polynomials} in the variables $x_1,\ldots,x_n$ over a field $K$ is the ring $K[x_j^{(i)}\mid i\geqslant 0,\, 1\leqslant j\leqslant n]$ with a derivation defined on the ring by $(x_j^{(i)})' := x_j^{(i + 1)}$. 
  This differential ring is denoted by $K\{x_1,\ldots,x_n\}$.
  \item An ideal $I$ of a differential ring $(R,{}')$ is called a {\em differential ideal} if, for all $a \in I$, $a' \in I$. For $F\subset R$, the smallest differential ideal containing the set $F$ is denoted by $[F]$.
    \item 
  For an ideal $I$ and element $a$ in a ring $R$, we denote \[I \colon a^\infty = \{r \in R \mid \exists \ell\colon a^\ell r \in I\}.\]
  This set is also an ideal in $R$.

  \item  An ideal $P$  of a commutative ring $R$ is said to be prime if, for all $a,b\in R$, if $ab\in P$ then $a \in P$ or $b\in P$.
    \item Given $\Sigma$ as in~\eqref{eq:ODEsystem2}, we define the differential ideal of $\Sigma$ as \[I_\Sigma=[Q(\bar{x}'-\bar{f}),Q(\bar{y}-\bar{g})]:Q^\infty \subset \mathbb{Q}(\bar{\alpha})\{\bar{x},\bar{y},\bar{u}\},\] where $Q$ is the common denominator of $\bar f$ and $\bar g$. By \cite[Lemma~3.2]{HOPY2020}, $I_\Sigma$ is a prime differential ideal.

 \item
  A {\em differential ranking} on $K\{x_1,\ldots,x_n\}$ is a total order $>$ on $X := \{x_j^{(i)}\mid i\geqslant 0,\, 1\leqslant j\leqslant n\}$ satisfying:
  \begin{itemize}
    \item for all $x \in X$, $x' > x$ and
    \item for all $x, y \in X$, if $x >y$, then $x' > y'$.
  \end{itemize}
  It can be shown that a differential ranking on $K\{x_1,\ldots,x_n\}$ is always a well order.   The ranking is \emph{orderly} if 
  moreover for all $i$, $j$,  $o_1$, and $o_2$, if $o_1 > o_2$, then $x_i^{(o_1)} > x_j^{(o_2)}$. 

\item
  For  $f \in K\{x_1,\ldots,x_n\} \backslash K$ and differential ranking $>$,
\begin{itemize}
    \item $\lead(f)$ is the element of $\{x_j^{(i)} \mid i \geqslant 0, 1 \leqslant j \leqslant n\}$ appearing in $f$ that is maximal with respect to $>$.
    \item The leading coefficient of $f$ considered as a polynomial in $\lead(f)$ is denoted by $\init(f)$ and called the initial of $f$. 
    \item The separant of $f$ is $\frac{\partial f}{\partial\lead(f)}$, the partial derivative of $f$ with respect to $\lead(f)$.
    \item The rank of $f$ is $\rank(f) = \lead(f)^{\deg_{\lead(f)}f}$.
    \item For $S \subset K\{x_1,\ldots,x_n\} \backslash K$, the set of initials and separants of $S$ is denoted by $H_S$.
    \item for $g \in K\{x_1,\ldots,x_n\} \backslash K$, say that $f < g$ if $\lead(f) < \lead(g)$ or $\lead(f) = \lead(g)$ and $\deg_{\lead(f)}f < \deg_{\lead(g)}g$.
\end{itemize}
    \item For $f, g \in K\{x_1,\ldots,x_n\} \backslash K$, $f$ is said to be reduced w.r.t. $g$ if no proper derivative of $\lead(g)$ appears in $f$ and $\deg_{\lead(g)}f <\deg_{\lead(g)}g$.
    \item 
    A subset $\mathcal{A}\subset K\{x_1,\ldots,x_n\} \backslash K$
    is called {\em autoreduced} if, for all $p \in \mathcal{A}$, $p$ is reduced w.r.t. every  element of $\mathcal A\setminus \{p\}$. 
    One can show that every autoreduced set has at most $n$ elements (like a triangular set but unlike a Gr\"obner basis in a polynomial ring).
    \item Let $\mathcal{A} = \{A_1, \ldots, A_r\}$ and $\mathcal{B} = \{B_1, \ldots, B_s\}$ be autoreduced sets such that $A_1 < \ldots < A_r$ and $B_1 < \ldots < B_s$. 
    We say that $\mathcal{A} < \mathcal{B}$ if
    \begin{itemize}
      \item $r > s$ and $\rank(A_i)=\rank(B_i)$, $1\leqslant i\leqslant s$, or
      \item there exists $q$ such that $\rank(A_q) <\rank(B_q)$ and, for all $i$, $1\leqslant i< q$, $\rank(A_i)=\rank(B_i)$.
    \end{itemize}
    \item An autoreduced subset of the smallest rank of a differential ideal $I\subset K\{x_1,\ldots,x_n\}$
    is called a {\em characteristic set} of $I$. One can show that every non-zero differential ideal in $K\{x_1,\ldots,x_n\}$ has a characteristic set. Note that a characteristic set does not necessarily generate the ideal.

\end{enumerate}

\begin{definition}[IO-identifiability]\label{def:IO-ident}
The smallest field $k$ such that
\begin{itemize}
\item $\mathbb{Q} \subset k \subset \mathbb{Q}(\bar{\alpha})$ and 
\item $I_\Sigma \cap \mathbb{Q}(\bar{\alpha})\{\bar{y},\bar{u}\}$ is generated as a differential ideal by $I_\Sigma \cap k\{\bar{y},\bar{u}\}$
\end{itemize}
is called \emph{the field of globally IO-identifiable functions}.  In differential algebra, such a field $k$ is called the field of definition of the ideal $I_\Sigma$ \cite[page~125]{Kolchin}.

We call $h \in \mathbb{Q}(\bar{\alpha})$  \emph{{globally} IO-identifiable} if $h \in k$. We also call $h \in \mathbb{Q}(\bar{\alpha})$  \emph{locally IO-identifiable} if $h$ is in the algebraic closure of the field $k$.  
\end{definition}

\begin{remark} By \cite[Theorem~19]{allident}, IO-identifiability is equivalent to multi-experimental identifiability \cite[Definition~16]{allident}. The latter is built using the notion of identifiability, which can be stated using the language of algebra~\cite[Definition~7]{allident} or in a more analytic way~\cite[Definition~2.5]{HOPY2020}.
\end{remark}

\begin{definition}[IO-equations]
Given a differential ranking on the differential variables $\bar y$ and $\bar u$, the \emph{IO-equations} are defined as the monic characteristic presentation of the prime differential ideal $I_\Sigma \cap \mathbb{Q}(\bar{\alpha})\{\bar{y},\bar{u}\}$ with respect to this ranking (see \cite[Definition~6 and Section~5.2]{ident-compare} for more details). For a given differential ranking, such a monic characteristic presentation is unique~\cite[Theorem~3]{Boulier2000}.
\end{definition}

Let $\bar\beta$ generate the field of globally IO-identifiable functions of the parameters. The tuple $\bar\beta$ can be computed as the set of coefficients of input-output equations, which are a canonical (still can depend on the choice of ranking on the variables) characteristic set of the projection of~\eqref{eq:ODEsystem2} to the   $(\bar u, \bar y)$-variables~\cite[Corollary~1]{ident-compare}. On a computer, this can be done, for instance, in {\sc Maple} using {\tt RosenfeldGroebner} or {\tt ThomasDecomposition}. An implementation that further simplifies $\bar\beta$ is available at \url{https://github.com/pogudingleb/AllIdentifiableFunctions} as a part of~\cite{allident}.

\section{Main algorithm}\label{sec:2} 
We break down our approach into the following several {\bf steps}, which we describe and justify in detail in Theorem~\ref{thm:main}:
\begin{enumerate}
\item Find input-output equations, view them as algebraic equations $E$, and compute the rational parametrization of the variety $V$ defined by $(E):H_E^\infty$ induced by the Lie derivatives of the output variables. 
\item Create a polynomial system of equations based on the computed parametrization 
 whose solutions
provide another rational parametrization of $V$ but now over (the algebraic closure of) the field of identifiable functions. Pick a solution, and therefore, a locally IO-identifiable rational parametrization of $V$.   Whenever it exists, pick such a solution that results in a globally IO-identifiable rational parametrization of $V$.
\item Reconstruct a locally (or globally if it exists) IO-identifiable ODE system from the new rational parametrization, cf.~\cite{Forsman}. 
\item By comparing the two rational parametrizations of $V$, find the corresponding change of state variables using Gr\"obner bases. 
\end{enumerate}
\begin{theorem}\label{thm:main} There is an algorithm solving the local IO-identifiable reparametrization problem from Section~\ref{sec:prob2} for system~\eqref{eq:ODEsystem2}, whose detailed steps are given in the proof. 
Furthermore, 
\begin{itemize}
\item as in \cite[Theorem~1]{OPPS2023},   if the sum of the orders with respect to the $
\bar y$-variables of the IO-equations is equal to the dimension of the model,
the state variables of the reparametrized system can be expressed as algebraic functions of $\bar x$ and $\bar\alpha$.
\item If the ODE system~\eqref{eq:ODEsystem2} has a globally IO-identifiable reparametrization whose Lie derivatives have monomial support being a subset of the monomial support of the Lie derivatives for~\eqref{eq:ODEsystem2}, then we can find this globally IO-identifiable reparametrization of~\eqref{eq:ODEsystem2}.
\end{itemize}
\end{theorem}
\begin{proof}
We follow the four steps outlined above.
\begin{enumerate}
\item 
{\bf Rational parametrization of IO-equations.}\label{seq:RPar}
By computing Lie derivatives of $\bar{y},\ldots,\bar{y}^{(n)}$ using~\eqref{eq:ODEsystem2}, for each $i$, we can write $y_s^{(i)}$ as a 
 rational function 
\begin{equation}\label{eq:param1}
\begin{multlined}
h_{s,i}\left(\hat{x}, \bar{\alpha}, \bar{u}, \ldots, \bar{u}^{(i)}\right) \\= \frac{\sum_{m\in M_1} m(\bar{\alpha})\cdot p_{s,m}\left(\hat{x}, \bar{u}, \ldots, \bar{u}^{(i)}\right)}
{  \sum_{m\in M_2} m(\bar{\alpha})\cdot q_{s,m}\left(\hat{x}, \bar{u}, \ldots, \bar{u}^{(i)}\right)}
\end{multlined}
\end{equation}
for some sets $M_1$ and $M_2$ of polynomials $m$ in the indeterminates $\bar\alpha$, where $\hat{x}$ are the variables from $\bar{x}$ that explicitly appear in the Lie derivatives for $\bar{y},\ldots,\bar{y}^{(n)}$, and the $p$'s and $q$'s are polynomials over $\mathbb{Q}$. 
Let 
\begin{equation}\label{eq:IO}
Y_s\left(\bar{\beta},\bar{y},\ldots,\bar{y}^{(n_s)},\bar{u},\ldots,\bar{u}^{(n_s)}\right)=0,\quad 1\leqslant s\leqslant |\bar y|
\end{equation}
be the input-output equations $E $ with respect to an {\em orderly} ranking on $\mathbb{Q} \{ \bar{y} \}$,  where here we write $|\bar y|$ for
the length of the tuple of variables $\bar y$. Note that the  rational functions
$h_{s,i}$ considered as functions from the affine space  $\mathbb{A}^{|\hat x|}$ with $\hat x$ coordinates to the affine $(n_1+\ldots+n_{|\bar y|})$-space is a unirational parametrization over $\mathbb{Q}(\bar \beta)\langle\bar u\rangle$, 
 $h:\mathbb{A}^{|\hat x|} \to V$, of the affine variety $V$ defined over $\mathbb{Q}(\bar \beta)\langle\bar u\rangle$ by the input-output equations: $V$ is the zero set of the ideal
\[ I_V := 
(E ):H_{E }^\infty.
\]
Since $I_\Sigma$ is a prime differential ideal, the differential ideal $I_\Sigma \cap \mathbb{Q}(\bar\beta)\{\bar y,\bar u\}$ is prime. Since  $E $ is a characteristic set of $I_\Sigma \cap \mathbb{Q}(\bar\beta)\{\bar y,\bar u\}$, we have
\[I_\Sigma \cap \mathbb{Q}(\bar\beta)\{\bar y,\bar u\} = [E ]:H_{E }^\infty.\]
By Rosenfeld's lemma from differential algebra \cite[Lemma~III.8.5]{Kolchin}, the polynomial ideal $I_V$ is prime as well, and so $V$ is an irreducible affine variety.

If $|\hat x| = \dim V$, define $\tilde x := \hat x$. If $|\hat x| > \dim V$, 
then we look for a linear change of variables $\hat{x} = A \tilde{x}$ for some matrix $A$ over $\mathbb{Q}$ of rank $\dim V$ defining
a linear map $L:\mathbb{A}^{\dim V} \to \mathbb{A}^{|\hat x|}$ so that 
$h \circ L : \mathbb{A}^{\dim V} \to V$ is a unirational parameterization of $V$.  On 
general grounds, almost any $A$ works.  Indeed, the condition is $\rank (dh\circ A) = \dim V$, where $dh$ is the differential of $h$.

We then have \begin{equation}\label{eq:equaldimensions}
n_1+\ldots+n_{|\bar{y}|}=\operatorname{trdeg} \mathbb{Q}(\bar\alpha)\langle \bar{y}, \bar{u}\rangle / \mathbb{Q}(\bar{\alpha})\langle \bar{u}\rangle = |\tilde x|.
\end{equation}
Let $M = \{m_1,\ldots,m_q\}$.
Note that 
\begin{equation}\label{eq:IOvanishing} Y_s\left(\bar{\beta},\bar{y},\ldots,\bar{y}^{(n_s)},\bar{u},\ldots,\bar{u}^{(n_s)}\right)|_{\bar{y}^{(i)} = h_{s,i},1\leqslant i\leqslant n_s}=0
\end{equation}
holds for all $s$, $1\leqslant s\leqslant |\bar y|$.
\item{\bf Rational parametrization over identifiable parameters.}
Consider the new indeterminates $z_1,\ldots,z_q$ and the 
rational functions
\[H_{s,i}(\bar{z},\tilde{x},\bar{u},\ldots,\bar{u}^{(i)}) := h_{s,i}|_{L(\tilde x) = \hat x,\ m_j=z_j, 1\leqslant j\leqslant q},\]
where  $0\leqslant i \leqslant n_s$.
Consider the system of polynomial equations (after clearing out the denominators)
\begin{equation}\label{eq:gamma}
Y_s(\bar{\beta},H_{s,0},\ldots,H_{s,n_s},\bar{u},\ldots,\bar{u}^{(n)})
=0
\end{equation}
in the variables $z_1,\ldots,z_q$. 
It follows from~\eqref{eq:IOvanishing} that the system has a solution in $\mathbb{Q}(\bar\alpha)$. Since the coefficients of the system belong to $\mathbb{Q}(\bar\beta)$, it has a solution $\bar{\gamma}$ in the algebraic closure of $\mathbb{Q}(\bar{\beta})$.
\item{\bf Identifiable ODE realization of the IO-equations given the new rational parametrization.}
Consider now \begin{equation}
\label{eq:newLie}H_{s,0}(\bar{\gamma},\bar{w},\bar{u}),\ldots,H_{s,n_s}(\bar{\gamma},\bar{w},\bar{u},\ldots,\bar{u}^{(n_s)}),
\end{equation}
in which we replaced $\tilde{x}$ by the new indeterminates $\bar{w}$,
and try to find an explicit ODE system (cf.~\cite{Forsman})
\begin{equation}\label{eq:reparam0}
\begin{cases}
\bar{w}' = F(\bar{\gamma},\bar{w},\bar{u},\ldots,\bar{u}^{(n+1)}),\\
\bar{y} = G(\bar{\gamma},\bar{w},\bar{u},\ldots,\bar{u}^{(n+1)} )
\end{cases}
\end{equation}
so that the input-output equations of~\eqref{eq:reparam0} coincide with~\eqref{eq:IO} as follows by making sure that~\eqref{eq:newLie} are the Lie derivatives of $\bar{y}$. We have 
\begin{align*}H_{1,1} &= \bar{y}' = H_{1,0}' =  \frac{\partial H_{1,0}}{\partial\bar{w}}\bar w'+\sum\frac{\partial H_{1,0}}{\partial\bar{u}}\bar{u}' \\ &= \frac{\partial H_{1,0}}{\partial\bar{w}}F+\frac{\partial H_{1,0}}{\partial\bar{u}}\bar{u}'\\
&\vdots\\
H_{1,n_1} &= y^{(n_1)}= H_{1,n_1-1}' \\&\quad= \frac{\partial H_{1,n_1-1}}{\partial\bar{w}}\bar w'+\sum\frac{\partial H_{1,n_1-1}}{\partial\bar{u}}\bar{u}' \\
&\quad =\frac{\partial H_{1,n_1-1}}{\partial\bar{w}}F+\sum_{i=0}^n\frac{\partial H_{1,n_1-1}}{\partial\bar{u}^{(i)}}\bar{u}^{(i+1)}\\
&\vdots
\end{align*}
Define an $(n_1+\ldots+n_{|\bar y|})$-vector
\[
H = \begin{pmatrix}
H_{1,1}-\sum_{i=0}^n\frac{\partial H_{1,0}}{\partial\bar{u}^{(i)}}\bar{u}^{(i+1)}\\
\vdots\\
H_{1,n_1}-\sum_{i=0}^n\frac{\partial H_{1,n_1-1}}{\partial\bar{u}^{(i)}}\bar{u}^{(i+1)}
\\
\vdots\\
H_{|\bar y|,1}-\sum_{i=0}^n\frac{\partial H_{|\bar y|,0}}{\partial\bar{u}^{(i)}}\bar{u}^{(i+1)}
\\
\vdots
\\
H_{|\bar y|,n_{|\bar y|}}-\sum_{i=0}^n\frac{\partial H_{|\bar y|,n_{|\bar y|}-1}}{\partial\bar{u}^{(i)}}\bar{u}^{(i+1)}
\end{pmatrix}
\]
and an $(n_1+\ldots+n_{|\bar y|})\times |\bar w|$-matrix (see~\eqref{eq:equaldimensions})
\begin{multline*}
dH =\\ \begin{pmatrix}\frac{\partial H_{1,0}}{\partial\bar{w}},
\ldots
,
\frac{\partial H_{1,n_1-1}}{\partial\bar{w}},
\ldots
,
\frac{\partial H_{|\bar y|,0}}{\partial\bar{w}},
\ldots
,
\frac{\partial H_{|\bar y|,n_1-1}}{\partial\bar{w}}
\end{pmatrix}^T.
\end{multline*}
Then the above translates into a linear system in $F$: \[dH\cdot F = H.\]
What if $\det dH$ is zero? Then go back and choose a different tuple $\bar\gamma$ satisfying~\eqref{eq:equaldimensions} and additionally  $\det dH \ne 0$. 
\item{\bf Corresponding change of variables.}
This step is done as \cite[Section~III, step 4]{OPPS2023}, which computationally is: solving the system of polynomial equations  (after clearing out the denominators) $H_{s,i} = h_{s,i}|_{L(\tilde x)=\hat x}$ for $\bar w$. This can be done, for instance, by doing a Gr\"obner basis computation with an elimination monomial ordering.
\end{enumerate}
\end{proof}
 
\begin{remark}\label{rem:2}
In all of our examples from Section~\ref{sec:examples}, it was possible to express the $\bar w$-variables as rational functions of the original state variables and parameters. However, for some examples, more general 
algebraic functions 
are required for this change of variables -- see Section~\ref{ex:noexprw}.
\end{remark}
 
\section{Explaining the approach using examples}\label{sec:examples}
In this section, we illustrate our approach using a series of examples, intentionally beginning with toy linear models to show the basics first. The non-linear examples are Lotka-Volterra models with input, a polynomial chemical reaction network model,  a rational (non-polynomial) biohydrogenation model, and a bilinear model with input. We  also include a linear compartmental model with input, for which the prior method of finding scaling identifiable reparametrizations failed but our more general method succeeded. We end the section with an example in which radicals appear in the change of variables.
\subsection{Turning local into global identifiability}
Consider the system
\begin{equation}\label{eq:1}
\begin{cases}
x_1' = ax_1,\\
x_2' = bx_2,\\
y = x_1+ x_2,
\end{cases}
\end{equation}
and so $\bar x = (x_1,x_2)$, $\bar y = y$, and $\bar \alpha = (a,b)$. There is no $\bar u$.
The input-output equation is
\begin{equation}\label{eq:IOeqexab}
y'' - (a+b) y' + ab\cdot y = 0.
\end{equation}
Therefore, $\bar\beta = (a+b, a\cdot b)$ and the identifiable functions are $K := \mathbb{Q}(a+b, a\cdot b)$, and so $a$ and $b$ are algebraic of degree 2 over $K$, therefore, are only locally identifiable. The approach from Section~\ref{sec:2} will proceed as follows. For $i = 0, 1, 2$, we will compute $y^{(i)}$ as a function $h_i(x_1,x_2, a, b)$:
\begin{equation}
\begin{aligned}\label{eq:Liederexab}
y &= h_0(x_1,x_2, a, b) = x_1 + x_2,\\
y' &= h_1(x_1,x_2, a, b) = x_1'+x_2' = ax_1 + bx_2,\\
y'' &=h_2(x_1,x_2, a, b)= x_1''+x_2'' = a^2x_1 + b^2x_2,
\end{aligned}
\end{equation}
and so $\hat x = (x_1,x_2)$, $\tilde x = \hat x$, and $M = \{1,a,b,a^2,b^2\}$. The equations~\eqref{eq:Liederexab} induce the following parametrization of the plane induced by~\eqref{eq:IOeqexab}, where, since the equation is linear, $(E ):H_E ^\infty = (E )$:
\begin{align}
Y_2 - (a&+b) Y_1 + ab\cdot Y_0 = 0,\label{eq:plane}\\
&\begin{aligned}
Y_0 &= x_1 + x_2,\\
Y_1 &= ax_1 + bx_2,\\
Y_2 &= a^2x_1 + b^2 x_2.
\end{aligned}\notag
\end{align}
We now define
\begin{equation}\label{eq:genparamexab}
\begin{aligned}
H_0 &= z_1w_1 + z_2w_2,\\
H_1&= z_3w_1 + z_4w_2,\\
H_2&= z_5w_1 + z_6w_2.
\end{aligned}
\end{equation}
and search for a reparametrization of~\eqref{eq:plane} of the form defined by~\eqref{eq:genparamexab}:
\[
(z_5w_1 + z_6w_2) - (a+b)(z_3w_1 + z_4w_2) + ab(z_1w_1 + z_2w_2) = 0,
\]
arriving at the following solution set in the $z$-variables:
\[
z_5 = -abz_1 + (a + b)z_3,\quad z_6 = -abz_2 + (a+ b)z_4.
\]
This solution set has $4$ free variables, $z_1,\ldots,z_4$. For the simplicity of the next steps, let us make the following choice:
\[ z_1 = 1,\ z_2 = 0,\ z_3 = 0,\ z_4=1,\]
which we can adjust later if necessary if the choice makes the next steps degenerate (a non-degenerate choice always exists according to Section~\ref{sec:2}). So, we have $z_5 = -ab$ and $z_6 = a+ b$, which turns~\eqref{eq:genparamexab} into
\begin{equation}\label{eq:genparamexabnew}
\begin{aligned}
H_0 &= w_1,\\
H_1&=w_2,\\
H_2&= -abw_1 + (a+b)w_2.
\end{aligned}
\end{equation}
We now construct an ODE realization of \eqref{eq:plane} from parametrization $Y_0 = H_0, Y_1 = H_1, Y_2 = H_2$ from~\eqref{eq:genparamexabnew} using the following equations:
\begin{align*}
w_1 &= H_0 = y,\\
w_1' &= H_0' = y' = H_1 = w_2,\\
w_2' &= H_1' = {(y')}' = y''= H_2 = -abw_1 + (a+b)w_2.
\end{align*}
Thus, we finally have
\[
\begin{cases}
w_1' = w_2,\\
w_2' = (a+b)w_2 - a b w_1,\\
y = w_1,
\end{cases}
\]
We now find the conversion from the $x$-variables to the $w$-variables:
\[
\begin{cases}
w_1 = H_0 = Y_0 =  x_1 + x_2,\\
w_2 = H_1 = Y_1 = ax_1 + bx_2.
\end{cases}
\]
\subsection{Making choices for the non-vanishing of $\det dH$}
Consider the system
\begin{equation}\label{eq:comparison}
\begin{cases}
x_1' = ax_2,\\
x_2' = bx_1,\\
y = x_1,
\end{cases}
\end{equation}
so $\bar x = (x_1,x_2)$, $\bar y = (y)$, $\bar\alpha = (a,b)$, and we have no $\bar u$.
The input-output equation is
\begin{equation}\label{eq:comparison:IO}
y'' - ab\cdot y = 0.
\end{equation}
Therefore, $\bar\beta = (a b)$ and $a b$ is globally identifiable but neither $a$ nor $b$ is identifiable. Following  the approach from Section~\ref{sec:2}, let us begin by computing Lie derivatives of $\bar y$. We have
\begin{equation}\label{eq:param:ex2}
\begin{aligned}
y &= h_0(x_1,x_2,a,b), = x_1, \\
y' &= h_1(x_1,x_2,a,b) =x_1' = ax_2,\\
y''&= h_2(x_1,x_2,a,b) = x_1'' = ax_2' = abx_1.
\end{aligned}
\end{equation} 
We have $\hat x = (x_1,x_2)$, $\tilde x = \hat x$, and $M=\{1,a,ab\}$. Equations~\eqref{eq:param:ex2} induce the following parametrization of the plane defined by the input-output equation, where, since the equation is linear, $(E ):H_E ^\infty = (E )$:
\begin{align}
&Y_2 -  ab\cdot Y_0 = 0,\label{eq:plane2}\\
&\begin{aligned}
Y_0 &= x_1,\\
Y_1 &= ax_2,\\
Y_2 &= abx_1.
\end{aligned}\notag
\end{align}
We now define
\begin{equation}\label{eq:genparamexab2}
\begin{aligned}
H_0 &= z_1w_1 ,\\
H_1&= z_2w_2,\\
H_2&= z_3w_1.
\end{aligned}
\end{equation}
and search for a reparametrization of~\eqref{eq:plane2} of the form defined by~\eqref{eq:genparamexab2}:
\[
z_3w_1 -  abz_1w_1 = 0,
\]
arriving at the following solution set in the $z$-variables: $z_3 = abz_1$.
This solution set has $2$ free variables, $z_1$ and $z_2$. For the simplicity of the next steps, let us make the following choice:
$z_1 = 1, z_2 = 0$,
which we can adjust later if necessary if the choice makes the next steps degenerate (a non-degenerate choice always exists according to Section~\ref{sec:2}). So, we have $z_3 = ab$, which turns~\eqref{eq:genparamexab2} into
\begin{equation}\label{eq:genparamexabnew2}
\begin{aligned}
H_0 &= w_1,\\
H_1&=0,\\
H_2&= abw_1.
\end{aligned}
\end{equation}
We now construct an ODE realization of \eqref{eq:plane2} from parametrization $Y_0 = H_0, Y_1 = H_1, Y_2 = H_2$ from~\eqref{eq:genparamexabnew2} using the following equations:
\begin{align*}
y &= H_0 = w_1,\\
w_1' &= H_0' = y' = H_1 = 0.
\end{align*}
However, we cannot find an ODE for $w_2$ because it does not appear in the $H_i$'s. So, let us instead choose a non-zero value for $z_2$, e.g., $z_2 = 1$. Then we have 
\begin{equation}\label{eq:genparamexabnew21}
\begin{aligned}
H_0 &= w_1,\\
H_1&=w_2,\\
H_2&= abw_1.
\end{aligned}
\end{equation}
and so we obtain:
\begin{align*}
y &= H_0 = w_1,\\
w_1' &= H_0' = y' = H_1 = w_2,\\
w_2' &= H_1' = {(y')}' = y''= H_2 = abw_1.
\end{align*}
Thus, we finally have
\[
\begin{cases}
w_1' = w_2,\\
w_2' = a b w_1,\\
y = w_1,
\end{cases}
\]
We now find the conversion from the $x$-variables to the $w$-variables:
\[
\begin{cases}
w_1 = H_0 = Y_0 =  x_1,\\
w_2 = H_1 = Y_1 = ax_2.
\end{cases}
\]

\subsection{Lotka-Volterra examples with input}
Consider the system
\begin{equation}
\label{eq:LV}
\begin{cases}
x_1' = ax_1 -bx_1x_2 + u,\\
x_2'= -cx_2 + dx_1x_2,\\
y = x_1
\end{cases}
\end{equation}
with two state variables $\bar{x} = (x_1,x_2)$, four parameters $\bar\alpha=(a,b,c,d)$, one output $\bar y = y$,  and one input $\bar u = u$. The input-output equation is
\begin{multline}
\label{eq:LVIO}
  yy'' - y'^2 - dy'y^2 + cyy' + uy'\\   + ady^3 + duy^2 - acy^2 - u'y - cuy=0.
\end{multline}
So, we have that the field of IO-identifiabile functions is
$\mathbb{Q}(d,c,ad,ac) = \mathbb{Q}(a,c,d)$.
A computation (in {\sc Maple}) shows that $(E ) = (E ):H_E ^\infty$ in this case.
The Lie derivatives of the $y$-variable are as follows:
\begin{equation}\label{eq:LieLV}
\begin{aligned}
y &= x_1, \\
y' &= -bx_1x_2 + ax_1 {  + u},\\
y''&= u' - bdx_1^2x_2 - bux_2 + au \\
&\quad+ \left(b^2x_2^2 + (-2a + c)bx_2 + a^2\right)x_1 
\end{aligned}
\end{equation} 
We then have $\tilde x =\hat x = (x_1,x_2)$, and 
we now define
\begin{equation}\label{eq:LieLVarb}
\begin{aligned}
H_0 &= z_1w_1\\
H_1 &=  z_2w_1w_2+z_3u+z_4w_1\\
H_2 &=
z_5w_1^2w_2+z_6w_1w_2^2+z_7uw_2+z_8w_1w_2\\
&  \quad+z_9u+z_{10}u'+z_{11}w_1
\end{aligned}
\end{equation}
Making the substitution $y=H_0, y' = H_1, y''=H_2$ into~\eqref{eq:LVIO}, we obtain the following polynomial system in  $z_1,\ldots,z_{11}$:
\begin{equation}\label{sys:LVupoly}
\begin{cases}
-dz_1^2z_2+z_1z_5 = 0\\
adz_1^3-dz_1^2z_4 = 0\\
z_1z_6-z_2^2 = 0\\
cz_1z_2+z_1z_8-2z_4z_2 = 0\\
-duz_1^2z_3-acz_1^2+duz_1^2+cz_1z_4\\\quad+z_1z_{11}-z_4^2 = 0\\
uz_1z_7-2uz_3z_2+uz_2 = 0\\
(cuz_3-cu+u'z_{10}-u')z_1\\
\quad+u(z_9-2z_3z_4+z_4) = 0\\
-u^2z_3^2+u^2z_3 = 0
\end{cases}
\end{equation}
In the above, $u$ and $u'$ are considered to be in the ground field for solving purposes, so these do not vanish. Also, if $z_1 = 0$, then~\eqref{eq:LieLVarb} is degenerate. So, we may assume that $z_1 \ne 0$. To preserve input, we may also assume $z_3 \ne 0$ (so, $z_3 = 1$). Solving system~\eqref{sys:LVupoly} in {\sc Maple} with these assumptions, we arrive at the following solution set, 
in which $z_1,z_7,z_{10}$  play the role of free variables:
\begin{multline*}
z_2 = z_1z_7, \ \ 
z_3 = 1,\ \ 
z_4 = az_1,\ \ 
z_5 = dz_1^2z_7,\ \  
z_6 = z_1z_7^2,\\ 
z_8 = (2a-c)z_1z_7,\ \ 
z_9 = \frac{au+(1-z_{10})u'}{u},\ \ 
z_{11} = a^2z_1.
\end{multline*}
Choosing (since, for us, it is sufficient to pick a solution) $z_1=z_7=z_{10} = 1$, we obtain
\[z_2 = z_3=z_6=1,\  z_4 =z_9= a,\ z_5 = d,\ z_8=2a-c,\ z_{11}=a^2.\]
Substituting into~\eqref{eq:LieLV}, we obtain
\[
\begin{aligned}
H_0 &= w_1, \\
H_1 &= (a+w_2)w_1+u,\\
H_2 &= dw_1^2w_2+\left(w_2^2+(2a-c)w_2+a^2\right)w_1+(a+w_2)u+u'.
\end{aligned}
\]
With the above, we now solve
\begin{gather*}
y= H_0=w_1,\\
w_1'=H_0',\\
(w_1w_2 + aw_1)' = H_1'=H_2.
\end{gather*}
and obtain
the following reparametrized system
\[
\begin{cases}
w_1' = aw_1+w_1w_2+u\\
w_2' = -cw_2+dw_1w_2
\end{cases}
\]
and to find the variable conversion, we solve the system
\begin{align*}
x_1 = h_0=H_0 &= w_1,\\
-bx_1x_2 + ax_1 + u= h_1= H_1 &= w_1w_2 + aw_1 + u,
\end{align*}
(we omitted the the equation with $H_2$ because the first two were already sufficient, and the additional one is too big to display and does not change the outcome) finding the following:
\begin{equation}\label{eq:LVchvar}
\begin{cases}
w_1 = x_1,\\
w_2 =  -bx_2.
\end{cases}
\end{equation}
 Here is another Lotka-Volterra model with input~\cite{LVinput}
\[
\begin{cases}
 x_1'=   ax_1 -bx_1x_2 + ux_1,\\
 x_2' = -cx_2 + dx_1x_2 + ux_2,\\
 y=x_1.
\end{cases}
\]
We omit the details because they are mostly the same as in the previous Lotka-Volterra model. The  globally  IO-identifiable parameters are $a,d,c$. According to our code, the same change of variables~\eqref{eq:LVchvar} results in the following globally IO-identifiable reparametrization:
\[
\begin{cases}
w_1' = aw_1+w_1w_2 +uw_1,\\
w_2' = -cw_2 + dw_1w_2+uw_2,\\
y=w_1.
\end{cases}
\]

\subsection{Chemical reaction network example}\label{sec:CRN}
Consider the following example based on~\cite[Example~5]{JPS2019}:
\[
\begin{cases}
x_1' = (2k_1 + k_4) x_2^2 - (k_2 + 2k_6) x_1^2 + (k_5-k_3)x_1x_2,\\
x_2'= -x_1',\\
y = x_1.
\end{cases}
\]
 Our calculation in {\sc Maple} shows that 
\begin{multline*}
 \cfrac{\left(4k_1 + k_3 + 2k_4 - k_5\right)^2}{2k_1+k_4},\\ \frac{(8k_2+16k_6)k_1 + (4k_2+8k_6)k_4 + (k_3-k_5)^2}{2k_1+k_4},
\end{multline*}
generate the field of globally IO-identifiable functions. 
 And, using our code, we obtain the following reparametrized  model equations with globally IO-identifiable parameters:
\begin{equation}\label{eq:CRNreparam}
\begin{cases}
w_1'=\tfrac{\left(4k_1 + k_3 + 2k_4- k_5\right)^2}{2k_1 + k_4}w_2^2\\
\qquad-\tfrac{(8k_2+16k_6)k_1 + (4k_2+8k_6)k_4 + (k_3-k_5)^2}{4(2k_1 + k_4)}w_1^2,\\
w_2'=\cfrac{w_1'}{2},\\
y=w_1
\end{cases}
\end{equation}
  and the following  
 linear change of variables resulting in~\eqref{eq:CRNreparam}:
\begin{equation}\label{eq:CRNchange}
\begin{cases}
w_1 = x_1,\\
w_2 = \cfrac{k_3 - k_5}{2(4k_1 + k_3 + 2k_4 - k_5)}x_1 \\ \qquad- \cfrac{2k_1 +k_4}{4k_1 + k_3 + 2k_4 - k_5}x_2.
\end{cases}
\end{equation}

\subsection{Biohydrogenation model}\label{sec:bioh}
Consider the following rational ODE model
\[
\begin{cases}
x_4' = - \cfrac{k_5  x_4}{k_6 + x_4},\\
x_5' = \cfrac{k_5x_4}{k_6 + x_4} - \cfrac{k_7 x_5}{k_8 + x_5 + x_6},\\
x_6' = \cfrac{k_7x_5}{k_8 + x_5 + x_6} - k_9 x_6\cfrac{ (k_{10} - x_6)}{k_{10}},\\
x_7'=k_9 x_6\cfrac{k_{10} - x_6}{k_{10}},\\
y_1 = x_4,\ \  y_2 = x_5,
\end{cases}
\]
(see \cite[system~(3), Supplementary Material 2]{MT2018}, initial conditions are assumed to be unknown, the choice of outputs is as in \url{https://maple.cloud/app/6509768948056064}).

We have $\bar x = (x_4,x_5,x_6,x_7)$, $\bar y = (y_1,y_2)$, $\bar\alpha = (k_5,k_6,k_7,k_8,k_9,k_{10})$, and there is no $\bar u$. 
Our {\sc Maple} code 
shows that the field of globally IO-identifiable functions is generated by
\[k_5,\ \  k_6,\ \  k_7,\ \  A := k_9^2,\ \ 
B := \frac{k_{10}}{k_9},\  \ C := k_9\cfrac{2k_8+k_{10}}{k_{10}}.
\]
We can see from this list that all parameters in this model $k_5,\ldots,k_{10}$ are at least locally IO-identifiable. Therefore, the approach from~\cite{OPPS2023} will leave this model as is, and so will not improve the identifiability properties of the model. In what follows, we will show how our approach makes the model globally IO-identifiable.
Our {\sc Maple} code then finds that
the resulting reparametrized system is
\begin{equation}\label{eq:biofinal}
\begin{cases}
w_1' = -k_5\cfrac{w_1}{k_6+w_1},\\
w_2' = \cfrac{((k_5 - k_7)w_2 + k_5w_3)w_1 - k_6k_7w_2}{(w_2 + w_3)(k_6 + w_1)},\\
w_3' = \frac{\frac{1}{B}w_2w_3^2 - Cw_2w_3 + \left(\frac{BC^2-AB}{4} + k_7\right)w_2 + \frac{1}{B}w_3^3 -Cw_3^2 + \frac{BC^2-AB}{4}w_3}{w_2+w_3},\\
y_1 = w_1, \ \ y_2 = w_2
\end{cases}
\end{equation}
under the following change of variables:
\[
\begin{cases}
w_1 = x_4,\\
w_2 = x_5,\\
w_3 = k_8+x_6.
\end{cases}
\]
Using SIAN~\cite{SIAN}, we have also checked to see that all parameters (and initial conditions) in~\eqref{eq:biofinal} are globally identifiable. The algorithm from~\cite{OPPS2023} cannot find this reparametrization because it has a different structure, e.g., a smaller number of state variables, among other things.
\subsection{Bilinear model with input}
Consider the model~\cite[Example~1]{LLV}:
\[
\begin{cases}
x_1'= - p_1x_1 + p_2u,\\
x_2'=  - p_3x_2 + p_4u,\\
x_3'=- (p_1+p_3)x_3 + (p_4x_1+p_2x_2)u,\\
y=x_3.
\end{cases}
\]
Our computation shows that the globally IO-identifiable functions are $p_1p_3, p_2p_4, p_1+p_3$ and that the following change of variables
\[
\begin{cases}
w_1 = p_2x_2+p_4x_1,\\
w_2 = -p_1p_2x_2-2p_1p_4x_1-2p_2p_3x_2-p_3p_4x_1,\\
w_3 = x_3
\end{cases}
\]
results in the following reparametrized globally IO-identifiable ODE system:
\[
\begin{cases}
w_1' = (p_1+p_3)w_1+2p_2p_4u+w_2,\\
w_2' = (-2p_1^2 - 5p_1p_3 - 2p_3^2)w_1 \\
\qquad- 3p_2p_4(p_1 + p_3)u + (-2p_1 - 2p_3)w_2,\\
w_3' = -(p_1+p_3)w_3+uw_1,\\
y = w_3
\end{cases}
\]
On the other hand, if one follows the algorithm from~\cite{OPPS2023}, one would arrive at the following system of equations and inequations in the unknowns $\widetilde{p_1},\widetilde{p_2},\widetilde{p_3},\widetilde{p_4}$: 
\[
\begin{cases}
p_1p_3 = \widetilde{p_1}\widetilde{p_3},\\
p_2p_4 = \widetilde{p_2}\widetilde{p_4},\\
p_1+p_3 = \widetilde{p_1}+\widetilde{p_3},\\
\widetilde{p_1}\widetilde{p_2}\widetilde{p_4}-\widetilde{p_2}\widetilde{p_3}\widetilde{p_4}\ne 0.

\end{cases}
\]
with solutions sought
over the algebraic closure of the field $\mathbb{Q}(p_1p_3,p_2p_4,p_1+p_3)$. This system does not have solutions over  $\mathbb{Q}(p_1p_3,p_2p_4,p_1+p_3)$ and the method from~\cite{OPPS2023} would just pick a value for $\widetilde{p_2}$, say, $\widetilde{p_2}=1$, and so $\widetilde{p_4}=p_2p_4$.  Thus the method from~\cite{OPPS2023} would arrive at the following ODE model, which is locally but not globally IO-identifiable:
\[
\begin{cases}
w_1'= - p_1w_1 + u,\\
w_2'=  - p_3w_2 + p_2p_4u,\\
w_3'=- (p_1+p_3)w_3 + (p_2p_4w_1+w_2)u,\\
y=w_3.
\end{cases}
\]
Here the limitation of~\cite{OPPS2023} that prevents the method from achieving global IO-identifiability is the requirement to keep the same monomial structure in each equation of the reparametrized vs. original ODE model, cf.~\cite[Section~IV.B]{OPPS2023}.
\subsection{Linear compartmental model with input}
We consider a model that does \textit{not} have an identifiable scaling reparametrization according to~\cite{MESHKAT201446} and thus could not be reparametrized using that approach.  We, however, are able to find a linear reparametrization using our approach.
\begin{equation}\label{eq:3rdlineareqs}
\begin{cases}
{x}_1' = a_{11} x_1 + a_{12} x_2 + u_1 \\
{x}_2' = a_{22} x_2 + a_{23} x_3 \\
{x}_3' = a_{31}x_1 + a_{32}x_2 + a_{33} x_3 \\
y_1 = x_1.
\end{cases}
\end{equation}
The IO-equation is
\begin{multline*}
y'''-(a_{11}+a_{22}+a_{33})y''+((a_{11}+a_{33})a_{22}+a_{11}a_{33}-a_{23}a_{32})y'\\+(-a_{11}a_{22}a_{33}+a_{11}a_{23}a_{32}-a_{12}a_{23}a_{31})y\\-u_1''+(a_{22}+a_{33})u_1'+(-a_{22}a_{33}+a_{23}a_{32})u_1=0.\end{multline*}
The coefficients of this equation generate the field of  globally IO-identifiable functions. After simplifying these generators using \url{https://github.com/pogudingleb/AllIdentifiableFunctions}, we obtain
\[a_{11},\ \ a_{12}a_{23}a_{31},\ \ a_{22} + a_{33},\ \ a_{22}a_{33} - a_{23}a_{32}\]
as generators of the field of globally IO-identifiable functions. 
To reparametrize~\eqref{eq:3rdlineareqs}, our next step is to find the Lie derivatives, which are:
\begin{gather*}
y = x_1,\\
y'= a_{11}x_1+a_{12}x_2+u_1,\\
y''=a_{11}^2x_1+a_{11}a_{12}x_2+a_{12}a_{22}x_2+a_{12}a_{23}x_3+u_1'+a_{11}u_1,\\
y'''=
(a_{11}^3+a_{12}a_{23}a_{31})x_1\\\qquad+(a_{11}^2a_{12}+a_{11}a_{12}a_{22}+a_{12}a_{22}^2+a_{12}a_{23}a_{32})x_2\\\qquad+(a_{11}a_{12}a_{23}+a_{12}a_{22}a_{23}+a_{12}a_{23}a_{33})x_3\\\qquad+u_1''+a_{11}u_1'+a_{11}^2u_1,
\end{gather*}
which, with undetermined coefficients, takes the form
\begin{equation}
\label{eq:H3rdlinout}
\begin{aligned}
H_0 &= w_1z_1,\\
H_1 &= z_2u_1+z_3w_1+z_4w_2,
\\
H_2 &=z_5u_1+z_6u_1'+z_7w_1+z_8w_2+z_9w_3,
\\
H_3 &= z_{10}u_1+z_{11}u_1'+z_{12}u_1''\\&\quad+z_{13}w_1+z_{14}w_2+z_{15}w_3.
\end{aligned}
\end{equation}
Since the IO-equation $E$ is linear, $(E) = I:H_E^\infty$, so we will be substituting the above $H$'s into $E$ to obtain the following system of linear equations in $z_1,\ldots,z_{15}$, which we solve and obtain
\begin{gather*}
z_{15} = (a_{11}+a_{22}+a_{33})z_9,\\
z_{14} = (a_{23}a_{32}-a_{11}a_{22}-a_{11}a_{33}-a_{22}a_{33})z_4\\ \qquad+(a_{11}+a_{22}+a_{33})z_8,\\
z_{13} = (a_{11}a_{22}a_{33}-a_{11}a_{23}a_{32}+a_{12}a_{23}a_{31})z_1\\ \qquad-(a_{11}a_{22}-a_{11}a_{33}-a_{22}a_{33}+a_{23}a_{32})z_3\\ \qquad+(a_{11}+a_{22}+a_{33})z_7\\
z_{12} = -\Big((a_{11}a_{22}+a_{11}a_{33}+a_{22}a_{33}-a_{23}a_{32})u_1z_2\\
\qquad
-(a_{11}+a_{22}+a_{33})(u_1z_5+u_1'z_6)+(a_{23}a_{32}-a_{22}a_{33})u_1\\
\qquad+(z_{11}+a_{22}+a_{33})u_1'+u_1z_{10}-u_1''\Big)/u_1'',
\end{gather*}
with $z_1,\ldots,z_{11}$ being free variables. We choose the following values for the free variables:
\begin{multline*}
z_1=z_2=z_4=z_6=z_9=z_{10}=z_{11}= 1,\\ z_3=z_5=z_8=a_{11},  \ \ z_7=a_{11}^2.
\end{multline*}
Substituting this entire solution in~\eqref{eq:H3rdlinout} and using the relationship $H_0' = H_1,\ H_1'=H_2,\  H_2'=H_3$, we obtain the following reparametrized system:
\[
\begin{cases}
w_1'= a_{11}w_1 + w_2 +u_1\\ 
w_2' = w_3 \\
w_3' = a_{12}a_{23}a_{31}w_1 + (a_{23}a_{32} - a_{22}a_{33})w_2 + (a_{22} + a_{33})w_3,\\
y_1=w_1.
\end{cases}
\]
We find the resulting (non-scaling) linear reparameterization:
\[ 
\begin{cases}
w_1 = x_1,\\   
w_2 = a_{12} x_2,\\
  w_3 = a_{12} a_{22} x_2 + a_{12} a_{23} x_3
  \end{cases}\]
  by setting equal~\eqref{eq:H3rdlinout} with the found $z$-values to the Lie derivatives and solving the resulting equations for $w_1,w_2,w_3$.
 
\subsection{Radicals in the change of variables}\label{ex:noexprw}
We now present an (artificially constructed) example of a rational ODE model for which, in our method, the expressions for $\bar w$ in terms of the original state variables and parameters involve radicals (see Remark~\ref{rem:2}). Consider the ODE model
\begin{equation}\label{eq:counter1}
\begin{cases}
x_1' = b^2 x_1 + ab (x_2^2 + x_2),\\
x_2' = \cfrac{x_1^2 - b (x_2^2 + x_2) -  b^2 (b^2  x_1 + ab (x_2^2 + x_2))}{ab (2  x_2 + 1)},\\
y = x_1.
\end{cases}
\end{equation}
The IO-equation of this system is
\[
ay''+y'-ay^2-b^2y=0.
\]
Therefore, the identifiable functions of the parameters are $\mathbb{Q}(a, b^2)$. The Lie derivatives are:
\begin{equation}\label{eq:Lie-counter}
\begin{aligned}
y &= x_1,\\
y' &= b^2x_1+ab(x_2^2+x_2),\\
y''&= x_1^2-b(x_2^2+x_2).
\end{aligned}
\end{equation}
To clarify, this example was manufactured by considering these Lie derivatives first:
\begin{align*}
y &= x_1,\\
y' &= b^2x_1+abx_2,\\
y''&= x_1^2-bx_2,
\end{align*}
and then replacing $x_2$ by $x_2^2 + x_2$ to obtain~\eqref{eq:Lie-counter}.
A computation in {\sc Maple} now shows that the following is a family of identifiable reparametrizations of~\eqref{eq:counter1}:
\[
\begin{cases}
w_1' = b^2w_1-\frac{z_2}{z_1}a(w_2^2+w_2),\\
w_2' = \cfrac{b^4z_1w_1-w_1^2z_1^2-z_2w_2(w_2+1)(ab^2+1)}{z_2a(2w_2+1)},
\end{cases}
\]
where $z_1$ and $z_2$ are any non-zero elements of $\mathbb{Q}(a,b^2)$, which originates from the following set of Lie derivatives
\begin{equation}\label{ex:counter-new-Lie}
\begin{aligned}
y  &= z_1w_1,\\
y' &=  z_1b^2w_1-z_2a(w_2^2+w_2)\\
y'' &= z_1^2w_1^2+z_2(w_2^2+w_2).
\end{aligned}
\end{equation}
Following our usual procedure to find the change of variables, we equate~\eqref{eq:Lie-counter} and~\eqref{ex:counter-new-Lie} and solve for $\bar w$ obtaining:
\[
\begin{cases}
w_1 = \cfrac{x_1}{z_1},\\
w_2 = \cfrac{-z_2 \pm \sqrt{z_2^2 - 4b(x_2^2+x_2)z_2}}{2z_2},
\end{cases}
\]
which stays radical for any non-zero choices of $z_1$ and $z_2$ from $\mathbb{Q}(a,b^2)$.

\section{Linear models}\label{sec:linear}
In this section, we focus on finding  globally IO-identifiable reparametrizations of linear ODE models. Theorem~\ref{thm:genexist} gives a general existence result of such reparametrizations based on analyzing our algorithm. Theorems~\ref{prop:oneout} and~\ref{prop:genmultout} provide explicit  globally IO-identifiable reparametrization formulas for linear compartmental models with single and multiple outputs, respectively. In these results, there are no inputs. However, they provide a basic approach that we then use to find identifiable reparametrizations for systems with inputs as follows.   Theorem~\ref{prop:onein} provides an explicit globally IO-identifiable reparametrization formula for a special case of linear compartmental models with a single input and output in the same compartment. Each of these explicit results is preceded by small examples that we calculated using our software and that gave a hint on what the general result should look like.
\subsection{General existence result}
\begin{theorem} \label{thm:genexist} Every model~\eqref{eq:ODEsystem2} in which $\bar f$ and $\bar g$ are linear has a globally  IO-identifiable linear reparametrization obtained by a linear change of variables. Moreover, this reparametrization can be found using the algorithm from Section~\ref{sec:2}.
\end{theorem}
\begin{proof}
Since $\bar f$ and $\bar g$ are linear, the IO-equations are linear in $\bar y,\bar y',\ldots, \bar y^{(n)}$, and so the corresponding variety $V$ is a hyperplane. The Lie derivatives of $\bar y$ are also linear in $\bar x$ (though could be non-linear in $\bar\alpha$, like in~\eqref{eq:Liederexab}). The embedding $L$ from step~\ref{seq:RPar} 
is linear. Since the coefficients of the monomials in the Lie derivatives are replaced by new indeterminates, the resulting system~\eqref{eq:gamma} is linear in the unknowns $z_1,\ldots,z_q$ (and is also consistent), and so it has a solution $\bar\gamma$ in $\mathbb{Q}(\bar\beta)$ itself (without taking the algebraic closure). Since $\mathbb{Q}(\bar\beta)$ is the field of globally IO-identifiable functions, $\bar\gamma$ is globally IO-identifiable.

With this solution $\bar\gamma$, the algorithm then proceeds to construct an ODE realization with the new Lie derivatives. This step is done by solving a consistent system of linear equations, and so the result is an ODE system with globally IO-identifiable parameters. Finally, the change of variables from the original $\bar x$ to the new $\bar w$ is linear as it can be found by setting the old and new expressions of the Lie derivatives of $\bar y, \bar y',\ldots,\bar y^{(n)}$, which are all linear (in $\bar x$ and $\bar w$, respectively).
\end{proof}

\subsection{Linear Compartmental Models}

\begin{definition}
Let $G$ be a directed graph with vertex set $V$ and set of 
directed edges $E$.  Each vertex $i \in V$ corresponds to a
compartment in our model and an edge $j \rightarrow i$ denotes 
a direct flow of material from compartment $j$ to
compartment $i$.  Also introduce three subsets of the vertices
$In, Out, Leak \subseteq V$ corresponding to the
set of input compartments, output compartments, and leak compartments
respectively.  To each edge $j \rightarrow i$, we associate
an independent parameter $a_{ij}$, the rate of flow
from compartment $j$ to compartment $i$.  
To each leak node $i \in Leak$, we associate an independent
parameter $a_{0i}$, the rate of flow from compartment $i$ leaving the system.

We associate a matrix $A$, called the \textit{compartmental matrix} to the graph and the set $Leak$
 in the following way:
\[
  A_{ij} = \left\{ 
  \begin{array}{l l l}
    -a_{0i}-\sum_{k: i \rightarrow k \in E}{a_{ki}} & \quad \text{if $i=j$ and } i \in Leak\\
        -\sum_{k: i \rightarrow k \in E}{a_{ki}} & \quad \text{if $i=j$ and } i \notin Leak\\
    a_{ij} & \quad \text{if $j\rightarrow{i}$ is an edge of $G$}\\
    0 & \quad \text{otherwise}\\
  \end{array} \right.
\]

Then we construct a system of linear ODEs with inputs and outputs 
as follows:
\begin{equation} \label{eq:main}
\dot{x}(t)=Ax(t)+u(t)  \quad \quad y_i(t)=x_i(t)  \mbox{ for } i \in Out
\end{equation}
 where $u_{i}(t) \equiv 0$ for $i \notin In$.
 The resulting model is called a   \textit{linear compartmental model}.
 \end{definition}

 For a model 
 as in~\eqref{eq:main}
 where there is a leak in every
compartment (i.e. $Leak = V$), it can greatly simplify the representation
to use the fact that the diagonal entries of $A$ are the
only places where the parameters $a_{0i}$ appear.
Since these are algebraically independent parameters,
we can introduce a new algebraically  independent 
parameter $a_{ii}$ for the diagonal entries (i.e. we make the substitution $a_{ii}  =  -a_{0i}-\sum_{k: i \rightarrow k \in E}{a_{ki}}$)
to get generic parameter values along the diagonal.
Identifiability questions in such a model are equivalent
to identifiability questions in the model with this reparametrized
matrix. 

We will be considering graphs that have some special connectedness properties.  We define these properties now.

\begin{definition} 
A \textit{path} from vertex $i_{k}$ to vertex $i_{0}$ in a directed graph $G$ is a sequence of 
vertices $i_{0},i_{1}, i_{2}, \ldots, i_{k}$
such that $i_{j+1} \to i_{j}$ is an edge for all $j = 0, \ldots, k-1$.
To a path $P = i_{0},i_{1}, i_{2}, \ldots, i_{k}$, we associate the
monomial $a^{P} = a_{i_{0}i_{1}}a_{i_{1}i_{2}}\cdots a_{i_{k-1}i_{k}}$,
which we refer to as a \textit{monomial path}.   
\end{definition}

\begin{definition}  
A directed graph $G$ is \textit{strongly connected} if there exists a directed path from each vertex to every other vertex.
\end{definition}

\subsection{Linear models without inputs}

In this section, we give a general technique to reparameterize a linear model without any inputs, but with one or more outputs.  The techniques we use in this section will be used to prove a special case for linear models with inputs.

\begin{example}
Consider the following model:
\begin{align*}
\dot{x}_1 &= a_{11} x_1 + a_{12} x_2 + a_{13} x_3 \\
\dot{x}_2 &= a_{21}x_1 + a_{22} x_2 + a_{23} x_3 \\
\dot{x}_3 &= a_{31}x_1 + a_{32}x_2 + a_{33} x_3 \\
y_1 &= x_1
\end{align*}
Here the identifiable functions are \begin{align*}
&a_{11}+a_{22}+a_{33},\\ &-a_{11}a_{22}-a_{11}a_{33}+a_{12}a_{21}+a_{13}a_{31}-a_{22}a_{33}+a_{23}a_{32},
\\ &a_{11}a_{22}a_{33}-a_{11}a_{23}a_{32}-a_{33}a_{12}a_{21}+a_{12}a_{23}a_{31}\\
&\qquad+a_{13}a_{32}a_{21}-a_{22}a_{13}a_{31}
\end{align*}
as these are the coefficients of the characteristic polynomial (up to sign).
Using the linear reparameterization:
\begin{align*}
X_1 &= x_1\\
X_2 &= a_{11}x_1 + a_{12} x_2 + a_{13} x_3 \\   X_3 &= (a_{11}^2 + a_{12}a_{21}+a_{13}a_{31})x_1\\
&\qquad+
(a_{11}a_{12}+a_{22}a_{12}+a_{13}a_{32})x_2
\\
&\qquad+(a_{11}a_{13}+a_{12}a_{23}+a_{13}a_{33})x_3 
\end{align*}
we obtain the following reparameterized system:
\begin{align*}
\dot{X}_1&=X_2 \\
\dot{X}_2& = X_3\\ 
    \dot{X}_3 &= (a_{11}a_{22}a_{33}-a_{11}a_{23}a_{32}-a_{33}a_{12}a_{21}+a_{12}a_{23}a_{31}\\&\qquad+a_{13}a_{32}a_{21}-a_{22}a_{13}a_{31})X_1 \\   &+(-a_{11}a_{22}-a_{11}a_{33}+a_{12}a_{21}+a_{13}a_{31}\\&\qquad-a_{22}a_{33}+a_{23}a_{32})X_2\\& + (a_{11}+a_{22}+a_{33})X_3
\end{align*}
\end{example}

\subsubsection{Reparametrization formula for linear systems with one output}
We will now derive an explicit formula for  globally IO-identifiable reparametrization of a linear ODE system with one output.
\begin{theorem}\label{prop:oneout} Consider a linear system over $\bar x = (x_1,\ldots,x_n)$: \[
    \begin{cases}
    \dot{\bar x}=A\bar x\\
    y_1 = x_1 = C\bar x,
    \end{cases}
    \]  where
    the graph corresponding to $A$ is strongly connected and 
    there is at least one leak and $C$ is the matrix where the $(1,1)$ entry is $1$ and all other entries are zero.
    Then using the linear reparameterization $\bar X=P \bar x$ given by:
\[
\begin{cases}
X_1=x_1\\
X_i = \sum\limits_{j=1} ^n p_{1j}^{(i-1)} x_j, & i=2,\ldots,n,
\end{cases}
\]where $p_{1j}^{(i-1)}$ is the sum of all monomial paths of length $i-1$ from $j$ to $1$, 
$$p_{1j}^{(i-1)}=\sum_{\operatorname{length}=i-1}  a_{1{k_1}}a_{{k_1}{k_2}}\cdots a_{{k_l}j}$$
we get a reparameterized   globally IO-identifiable (and, by \cite[Theorem~1]{OPT19}  globally identifiable as well) ODE system:
\begin{equation}\label{eq:singleoutreparam}
\begin{aligned}
\dot{X}_1&=X_2\\
\dot{X}_2&=X_3\\
&\vdots\\
\dot{X}_{n-1}&=X_n\\
\dot{X}_n&=-c_0 X_1 - c_1 X_2 - \ldots -c_{n-1} X_n
\end{aligned}
\end{equation}
where $c_i$ is the $n-i^{th}$ coefficient of the characteristic polynomial of $A$, $i=1,\ldots,n$.
The matrix $P$ is the $n\times n$ observability matrix:
\begin{align*}
\begin{pmatrix}
 C  \\
 CA \\
 \vdots \\
 C A^{n-1}
\end{pmatrix}
\end{align*}
\end{theorem}

\begin{proof} 
 A direct calculation shows that
\begin{equation}\label{eq:IOsingle}
y_1 ^{(n)} + c_{n-1} y_1^{(n-1)} + \ldots + c_1 \dot y_1 + c_0 y_1 = 0
\end{equation}
is the IO-equation of~\eqref{eq:singleoutreparam}.  This can be shown using the Laplace Transform/Transfer Function approach (see \cite{distefano-book} for more details).  This input-output equation is irreducible and of minimal order by  \cite[Theorem~3]{MeshkatSullivantEisenberg}. 
Notice the reparameterized system can be factored as:
$$
\dot{\bar X}=
\begin{pmatrix}
0 & 1 & 0 & 0 & \cdots & 0 \\
0 & 0 & 1 & 0 & \cdots & 0 \\
\vdots & \vdots & \vdots & \vdots & \vdots & \vdots \\
-c_0 & -c_1 & -c_2 & -c_3 & \cdots & -c_{n-1}
\end{pmatrix}
\bar X = \tilde{A}\bar X.$$ 
This is a standard result from differential equations on converting an $n^{th}$ order linear ODE (i.e. the input-output equation~\eqref{eq:IOsingle}):
into a first order system of $n$ ordinary differential equations via the procedure: 
\begin{align*}
X_1 &= y_1\\
X_2 &= \dot y_1  = \dot{X_1}\\
X_3 &= \ddot y_1  = \dot{X}_2\\ 
&\vdots\\
\dot{X}_n  &= y_1 ^{(n)} = -c_{n-1} y_1^{(n-1)} - \ldots - c_1 \dot y_1  - c_0 y_1 \\&= -c_0 X_1 - c_1 X_2 - \ldots -c_{n-1} X_n
\end{align*}
Now we show that this procedure leads to the linear reparameterization $\bar X=P\bar x$ given above.  We have: \begin{gather*}
X_1=y_1=x_1 = C\bar x\\
X_2= \dot{X}_1 = \dot{x}_1 = a_{11}x_1 + a_{12}x_2 + \ldots +a_{1n}x_n = CA \bar x\\
\begin{multlined}
X_3 = \dot{X}_2 = \ddot{X}_1 = \ddot{x}_1 = a_{11}\dot{x}_1+ a_{12}\dot{x}_2+\ldots+a_{1n}\dot{x}_n \\= a_{11} (a_{11}x_1+a_{12}x_2+\ldots+a_{1n}x_n)\\+ a_{12}(a_{21}x_1+a_{22}x_2+\ldots+a_{2n}x_n) + \ldots   
\\+a_{1n}(a_{n1}x_1+a_{n2}x_2+\ldots+a_{nn}x_n) = CA^2 \bar x
\end{multlined}
\\
\begin{multlined}
X_4 = \dot{X}_3 = \ddot{X}_2 = \dddot{X}_1 = \dddot{x}_1 = a_{11}\ddot{x}_1+ a_{12}\ddot{x}_2+\ldots+a_{1n}\ddot{x}_n \\= a_{11} (a_{11}\dot{x}_1+a_{12}\dot{x}_2+\ldots+a_{1n}\dot{x}_n)\\+ a_{12}(a_{21}\dot{x}_1+a_{22}\dot{x}_2+\ldots+a_{2n}\dot{x}_n) + \ldots  
\\+a_{1n}(a_{n1}\dot{x}_1+a_{n2}\dot{x}_2+\ldots+a_{nn}\dot{x}_n) = CA^3 \bar x
\end{multlined}
\\
\vdots \\
X_n = CA^{n-1} \bar x
\end{gather*}
Thus the matrix $P$ is the $n$ by $n$ observability matrix:
\begin{align*}
\begin{pmatrix}
 C  \\
 CA \\
 \vdots \\
 C A^{n-1}
\end{pmatrix}
\end{align*}
We can write $\dot{x}_1,\ldots,\dot{x}_n$ in terms of paths:
$$\dot{x}_i= \sum_{j=1} ^n p_{ij}^{(1)} x_j$$
for $i=1,\ldots,n$, $p_{ij}^{(1)}$ is the monomial path of length $1$ from $j$ to $i$.   
Then $\ddot{x}_i = \sum_{j=1} ^n p_{ij}^{(1)} \dot{x}_j$, which is:
$$\ddot{x}_i=\sum_{j=1} ^n p_{ij}^{(1)} \sum_{k=1} ^n p_{jk}^{(1)} x_k $$
which works out to:
$\ddot{x}_i=\sum_{j=1} ^n p_{ij}^{(2)} x_j $
where $p_{ij}^{(2)}$ is the sum of all monomial paths of length $2$ from $j$ to $i$,  
$$p_{ij}^{(2)}=\sum_{\operatorname{length}=2}  a_{i{k_1}}a_{{k_1}j}$$
We have that $x^{(n)}_i = \sum_{j=1} ^n p_{ij}^{(1)} x^{(n-1)}_j$ (by linearity) and now assume that:
$$x^{(n-1)}_i=\sum_{j=1} ^n p_{ij}^{(n-1)} x_j $$
where $p_{ij}^{(n-1)}$ is the sum of all monomial paths of length $n-1$ from $j$ to $i$. 
Then 
$$x^{(n)}_i=\sum_{j=1} ^n p_{ij}^{(1)} \sum_{k=1} ^n p_{jk}^{(n-1)} x_k $$
which works out to:
$x^{(n)}_i=\sum_{j=1} ^n p_{ij}^{(n)} x_j $,
where $p_{ij}^{(n)}$ is the sum of all monomial paths of length $n$ from $j$ to $i$, 
\[p_{ij}^{(n)}=\sum_{\operatorname{length}=n}  a_{1{k_1}}a_{{k_1}{k_2}}\ldots a_{{k_l}j}.\]
\end{proof}

\begin{corollary}
    The reparametrization in Theorem \ref{prop:oneout} yields $X_i$ that are linearly independent (in particular, are not zero) for $i=1,\ldots,n$. 
\end{corollary}

\begin{proof}
To show linear independence of the $X_i$, 
it is sufficient to show that
the Jacobian of the linear reparametrization is generically full rank.  
The 
Jacobian is given by the matrix $P$. 
This is the observability matrix:
\begin{align*}
\begin{pmatrix}
 C  \\
 CA \\
 \vdots \\
 C A^{n-1}
\end{pmatrix}
\end{align*}
where $C$ has $(1,1)$ entry equal to $1$, all others zero.  A $n-$compartment model is structurally observable if and only if the rank of the observability matrix is $n$ \cite{kalman}.  From  \cite[Theorem~1]{godfrey-chapman}, 
a compartmental model is structurally observable if and only if it is output connectable, which means there exists a path from every vertex to the output. Since $G$ is strongly connected by assumption, it is thus output connectable and this structurally observable, so the rank of the observability matrix is $n$.
\end{proof}

\subsubsection{Reparametrization formula for linear systems with multiple outputs}

\begin{example}\label{ex:2} 
Consider the following model:
\begin{align*}
\dot{x}_1 &= a_{11} x_1 + a_{12} x_2 + a_{13} x_3 + a_{14} x_4 + a_{15} x_5\\
\dot{x}_2 &= a_{21}x_1 + a_{22} x_2 + a_{23} x_3 + a_{24} x_4 + a_{25} x_5\\
\dot{x}_3 &= a_{31}x_1 + a_{32}x_2 + a_{33} x_3 + a_{34} x_4 + a_{35} x_5\\
\dot{x}_4 &= a_{41}x_1 + a_{42}x_2 + a_{43} x_3 + a_{44} x_4 + a_{45} x_5\\
\dot{x}_5 &= a_{51}x_1 + a_{52}x_2 + a_{53} x_3 + a_{54} x_4 + a_{55} x_5\\
y_1 &= x_1 \\
y_2 &= x_2
\end{align*}
Using the linear reparameterization:
\begin{gather*}
X_1 = x_1\\
X_2 = x_2
\\X_3 = \dot{x}_1 = a_{11} x_1 + a_{12} x_2 + a_{13} x_3 + a_{14} x_4 + a_{15} x_5 \\ X_4 = \dot{x}_2 = a_{21}x_1 + a_{22} x_2 + a_{23} x_3 + a_{24} x_4 + a_{25} x_5\\ X_5 =  \ddot{x}_1 = a_{11} \dot{x}_1 + a_{12} \dot{x}_2 + a_{13} \dot{x}_3 + a_{14} \dot{x}_4 + a_{15} \dot{x}_5 = \ldots
\end{gather*}
we obtain the following reparameterized system:
\begin{gather*}
\dot{X}_1=X_3 \\ \dot{X}_2 = X_4 \\ \dot{X}_3 = X_5 \\ \dot{X}_4 = \ddot{X}_2 \\ \dot{X_5} = \dddot{X}_1
\end{gather*}
\end{example}

The expressions for $\ddot{X}_2$ and $\dddot{X}_1$ on the right-hand side can then be written in terms of $X_1,...,X_5$, but we do not include these as the expressions get too big to fit on a page.

We can now generalize to the case of multiple outputs and write a reparametrized linear system.

\begin{theorem}\label{prop:genmultout}
    Consider
    \begin{itemize}
    \item a linear system over $\bar x = (x_1,\ldots,x_n)$: \[
    \begin{cases}
    \dot{\bar x}=A\bar x\\
    y_i = x_i, \quad i=1,\ldots,m
    \end{cases}
    \]
    \item $C$ the diagonal $m\times n$ matrix in which the $(i,i)$ entry is $1$ for $i=1,\ldots,m$, all other entries are zero
    \item the matrix $P$  given by the first $n$ rows of the observability matrix:
\begin{align} \label{eq:observ}
\begin{pmatrix}
 C  \\
 CA \\
 \vdots \\
 C A^{n-1}
\end{pmatrix}
\end{align}
    \end{itemize}
If the matrix $P$ is invertible, then, using the linear reparameterization $\bar X=P\bar x$,
we obtain a globally  IO-identifiable reparametrized ODE system
\[\dot{\bar X} = P A P^{-1} \bar X.\]
\end{theorem}
\begin{remark}
In coordinates, the new variables are
given by 
    \begin{equation} \label{eq:reparam}
\begin{aligned}
X_1&=x_1,\ldots,
X_m=x_m\\
X_{m+1}&=\dot{X}_1=\dot{x}_1\\
&\ \ \vdots\\
X_{2m}&=\dot{X}_{m}=\dot{x}_m\\
X_{2m+1}&=\dot{X}_{m+1}=\ddot{X}_1=\ddot{x}_1\\
&\ \ \vdots\\
X_n&=\dot{X}_{n-m}=\ddot{X}_{n-2m}=\ldots\\&=X^{(k)}_{n-km}=x_{n-km}^{(k)},
\end{aligned}
\end{equation}
    where $k\geqslant 0$ is an integer such that $m\geqslant n-km>0$. The reparameterized globally IO-identifiable ODE system is: 
        \begin{equation} \label{eq:reparamODE}\begin{cases}\dot{X}_1=X_{m+1},\\ \dot{X}_2=X_{m+2},\\
        \qquad\vdots
 \\  
\dot{X}_{m+1}=X_{2m+1}\\
  \qquad \vdots\\
   \dot{X}_{n-m}=X_n,\dot{X}_{n-m+1}=\ddot{X}_{n-m+1-m}\\
   \qquad\vdots\\
  \dot{X}_{n}=\ddot{X}_{n-m}
  \end{cases}
  \end{equation}
\end{remark}
\begin{remark} It would be interesting to know for what classes of linear systems, the matrix $P$ is invertible. For instance, it is invertible in Example~\ref{ex:2}. On the other hand, if $A$ is the zero matrix and $m < n$, then $P$ is not invertible. 
\end{remark}
\begin{proof}
Note that we trivially set $X_1=x_1,\ldots,X_m=x_m$ as we do not want to change the outputs. Following this same technique as in Theorem \ref{prop:oneout}, we can describe our reparametrization~\eqref{eq:reparam} in terms of $C$ and powers of $A$.  We have that \begin{gather*}(X_1,\ldots,X_m)=C \bar x, \\(X_{m+1},\ldots,X_{2m}) = CA \bar x, \ldots, (X_{n-m},\ldots,X_n)=CA^k \bar x.
\end{gather*}
This gives the first $n$ rows of the observability matrix in \eqref{eq:observ}.   Our reparametrization in~\eqref{eq:reparam} gives the right-hand-side expressions for $\dot{X}_1,\ldots,\dot{X}_{n-m}$, etc, in \eqref{eq:reparamODE} by setting them equal to $X_{m+1},\ldots,X_n$ until all variables $X_i$ have been exhausted. The expressions for $\dot{X}_{n-m+1},\ldots,\dot{X}_n$ in~\eqref{eq:reparamODE} can be obtained by taking derivatives of the first $n-m$ equations in~\eqref{eq:reparamODE}, as the variables $X_{n-m+1},\ldots,X_n$ appear in the first $n-m$ equations on the right-hand side of~\eqref{eq:reparamODE} since $n-m+1<n$ for $m>1$.  
This introduces second order derivatives (and higher) of the variables $X_1,\ldots,X_n$. 
To get the precise form of the right-hand-side of \eqref{eq:reparamODE} in terms of $\bar X$, 
we use the linear reparametrization $\bar X=P \bar x$ and get the reparametrized ODE system $\dot{\bar X}=PAP^{-1}\bar X$.

What's only left to prove is that this reparametrized ODE system is, in fact, a  globally  IO-identifiable reparametrization. Note that each of the expressions $X_1,\ldots,X_n$ can themselves be written in terms of $y_1,\ldots,y_m$ or their derivatives as follows: 
\[\begin{cases}
X_i=x_i=y_i,& i=1,\ldots,m,\\
X_j=\dot{X}_{j-m}=\dot y_{j-m},& j=m+1,\ldots,2m,\\ X_k=\dot{X}_{k-m}=\ddot y_{k-2m},& k=2m+1,\ldots,3m,\\
\quad\vdots
\end{cases}\]
 Thus, we can use  substitution to obtain that the equations in $\dot{X}_{n-m+1},\ldots,\dot{X}_n$ in~\eqref{eq:reparamODE} are $m$ IO-equations. As the coefficients of the IO-equations are  globally  IO-identifiable by definition, we have a   globally  IO-identifiable reparametrization.
\end{proof}

\begin{remark} Theorem~\ref{prop:oneout} can be seen as a special case of Theorem~\ref{prop:genmultout} for the case where $m=1$, and we thus get a single input-output equation in our identifiable reparametrization.
\end{remark}

\subsection{Linear models with inputs}

Finding an identifiable reparametrization in the case of linear models with inputs is a trickier problem than without inputs.  In \cite{MESHKAT201446}, necessary and sufficient conditions were found for identifiable scaling reparametrizations, but no work was done on the general case of linear reparametrizations (i.e. not just scaling).  We generalize our result for linear reparametrization without input to consider the special case of models with a single input and single output in the first compartment and have a single incoming edge and single outgoing edge to compartment $1$, which we call a single-in-single-out matrix.  

\begin{definition}
    A matrix $A$ is called a single-in-single-out matrix if $A$ is strongly connected and the entries of the matrix $A$ satisfy the following conditions: \[ 
    \begin{cases}
        a_{1i} \neq 0 \text{ for some unique } i \\
        a_{1j} = 0 \text{ for all } j \neq i\\
        a_{k1} \neq 0 \text{ for some unique } k \neq i \\
        a_{l1} = 0 \text{ for all } l \neq k \\
            \end{cases}
            \]
and there are no cycles of length $i$ for $i=2, \ldots, n-1$ involving compartment $1$.
\end{definition}

\begin{theorem}\label{prop:onein} Consider a linear system over $\bar x = (x_1,\ldots,x_n)$: \[
    \begin{cases}
    \dot{\bar x}=A\bar x + B\bar u\\
    y_1 = x_1 =  C\bar x,
    \end{cases}
    \]  where
    the graph corresponding to $A$ is a single-in-single-out matrix and 
    there is at least one leak and $B$ and $C$ are the matrices where the $(1,1)$ entry is $1$ and all other entries are zero.
    Then using the linear reparameterization $\bar X=P \bar x$ given by:
\[
\begin{cases}
X_1=x_1\\
X_i = \sum\limits_{j=1} ^n p_{1j}^{(i-1)} x_j, & i=2,\ldots,n,
\end{cases}
\]where $p_{1j}^{(i-1)}$ is the sum of all monomial paths of length $i-1$ from $j$ to $1$, 
\[p_{1j}^{(i-1)}=\sum_{\operatorname{length}=i-1}  a_{1{k_1}}a_{{k_1}{k_2}}\cdots a_{{k_l}j}\]
we get a reparameterized  {  globally} IO-identifiable (and, by \cite[Theorem~1]{OPT19} {  globally} identifiable as well) ODE system:
\begin{equation}\label{eq:singleoutreparam}
\begin{aligned}
\dot{X}_1&=X_2 + u\\
\dot{X}_2&=X_3 + a_{11} u\\
&\vdots\\
\dot{X}_{n-1}&=X_n + a_{11}^{n-2} u\\
\dot{X}_n&=-c_0 X_1 - c_1 X_2 - \ldots -c_{n-1} X_n + a_{11}^{n-1} u
\end{aligned}
\end{equation}
where $c_i$ is the $n-i^{th}$ coefficient of the characteristic polynomial of $A$, $i=1,\ldots,n$.
The matrix $P$ is the $n\times n$ observability matrix:
\begin{align*}
\begin{pmatrix}
 C  \\
 CA \\
 \vdots \\
 C A^{n-1}
\end{pmatrix}
\end{align*}
\end{theorem}

\begin{proof}
    The IO-equation of our system 
    is
\begin{multline}\label{eq:IOsingle}
y_1 ^{(n)} + c_{n-1} y_1^{(n-1)} + \ldots + c_1 \dot y_1 + c_0 y_1 = \\
u_1 ^{(n-1)} + d_{n-2} u_1^{(n-2)} + \ldots + d_1 \dot u_1 + d_0 u_1,
\end{multline}
 where $c_0, c_1, \ldots , c_{n-1}$ are the $n$ coefficients of the characteristic polynomial of $A$ and $d_0, d_1, \ldots , d_{n-2}$ are the $n-1$ coefficients of the characteristic polynomial of $A_{11}$, which the is the matrix $A$ with the first row and first column removed and is derived in \cite{MESHKAT201446}.  This can also be shown using the Laplace Transform/Transfer Function approach.  This input-output equation is irreducible and of minimal order by  \cite[Theorem~3]{MeshkatSullivantEisenberg}. 

By extending the standard technique of lifting an $n$th order ODE~\eqref{eq:IOsingle} to a system with a companion matrix, for a general matrix $A$, the reparameterized system can be factored as:
\[
\dot{\bar X}=
\begin{pmatrix}
0 & 1 & 0 & \cdots & 0 \\
0 & 0 & 1 & \cdots & 0 \\
\vdots & \vdots & \vdots & \vdots & \vdots \\
-c_0 & -c_1 & -c_2 & \cdots & -c_{n-1}
\end{pmatrix}
\bar X +
\begin{pmatrix}
1 \\
p_{11}^{(1)} \\
p_{11}^{(2)} \\
\vdots \\
p_{11}^{(n-1)}
\end{pmatrix}
u_1.\]
However, the terms $p_{11}^{(1)}, \ldots, p_{11}^{(n-1)}$ are not globally identifiable in general.  Recall that $p_{11}^{(i)}$ is the sum of all monomial paths of length $i$ from $1$ to $1$.  If $A$ is a single-in-single-out matrix, then $p_{11}^{(i)}$ reduces to just $a_{11}^{i}$  
\[
\dot{\bar X}=
\begin{pmatrix}
0 & 1 & 0 & \cdots & 0 \\
0 & 0 & 1 & \cdots & 0 \\
\vdots & \vdots & \vdots & \vdots & \vdots \\
-c_0 & -c_1 & -c_2 & \cdots & -c_{n-1}
\end{pmatrix}
\bar X +
\begin{pmatrix}
1 \\
a_{11} \\
a_{11}^2 \\
\vdots \\
a_{11}^{n-1}
\end{pmatrix}
u_1.\]
We note that the parameter $a_{11}$ is globally identifiable as it is equal to $-c_{n-1} + d_{n-2}$, i.e. $\tr(A)-\tr(A_{11})$.

The rest of the proof is identical to the proof of Theorem~\ref{prop:oneout} but with the addition of $u_1$ as follows:
\begin{align*}
X_1 &= y_1\\
X_2 + u_1 &= \dot y_1  = \dot{X_1}\\
X_3 + a_{11} u_1 &= \ddot y_1  = \dot{X}_2\\ 
&\vdots\\
\dot{X}_n + a_{11}^{n-1} u_1  &= y_1 ^{(n)} = -c_{n-1} y_1^{(n-1)} - \ldots - c_1 \dot y_1  - c_0 y_1 \\&= -c_0 X_1 - c_1 X_2 - \ldots -c_{n-1} X_n
\end{align*}
We now have:
\begin{gather*}
X_1=y_1=x_1 = C\bar x,\\
\begin{multlined}
X_2 = \dot{X}_1 - u_1 = \dot{x}_1 - u_1 \\= a_{11}x_1 + a_{12}x_2 + \ldots +a_{1n}x_n = CA \bar x,
\end{multlined}\\
\begin{multlined}
X_3  = \dot{X}_2 - a_{11} u_1= \ddot{X}_1 -a_{11} u_1 - \dot u_1= \ddot{x}_1 -a_{11} u_1 -\dot u_1 \\ = a_{11}\dot{x}_1+ a_{12}\dot{x}_2+\ldots+a_{1n}\dot{x}_n - a_{11}u_1\\= a_{11} (a_{11}x_1+a_{12}x_2+\ldots+a_{1n}x_n + u_1)\\+ a_{12}(a_{21}x_1+a_{22}x_2+\ldots+a_{2n}x_n) + \ldots   
\\+a_{1n}(a_{n1}x_1+a_{n2}x_2+\ldots+a_{nn}x_n) - a_{11}u_1= CA^2 \bar x,
\end{multlined}
\\
\vdots \\
X_n = CA^{n-1} \bar x
\end{gather*}    
\end{proof}

\begin{example}
Consider the following model:
\begin{align*}
\dot{x}_1 &= a_{11} x_1 + a_{13} x_3 +u_1 \\
\dot{x}_2 &= a_{21}x_1 + a_{22} x_2 + a_{23} x_3 \\
\dot{x}_3 &= a_{32}x_2 + a_{33} x_3 \\
y_1 &= x_1
\end{align*}
Notice that $a_{12}$ and $a_{31}$ have been set to zero to satisfy the single-in-single-out condition. This is an example of a model for which there is no identifiable scaling reparametrization using the results from \cite{MESHKAT201446} (it is not an identifiable cycle model), thus, by Theorem~\ref{thm:genexist}, a linear reparametrization must be made.  Linear compartmental models like this one come up in areas such as pharmacokinetics, physiology, cell biology, and ecology \cite{MeshkatSullivantEisenberg}.
The identifiable functions are \begin{align*}
&a_{11},
\\ &a_{13}a_{32}a_{21}, \\
& a_{22} + a_{33}, \\
& a_{22}a_{33}-a_{23}a_{32},
\end{align*}
which can be easily found from the coefficients of the input-output equation.
Using the linear reparameterization:
\begin{align*}
X_1 &= x_1\\
X_2 &= a_{11}x_1 + a_{12} x_2 + a_{13} x_3 \\   X_3 &= (a_{11}^2)x_1+
(a_{13}a_{32})x_2
+(a_{11}a_{13}+a_{13}a_{33})x_3 
\end{align*}
we obtain the following reparameterized system:
\begin{align*}
\dot{X}_1&=X_2 + u_1\\
\dot{X}_2& = X_3 + a_{11} u_1\\ 
    \dot{X}_3 &= (a_{11}a_{22}a_{33}-a_{11}a_{23}a_{32}+a_{13}a_{32}a_{21})X_1 \\   &+(-a_{11}a_{22}-a_{11}a_{33}-a_{22}a_{33}+a_{23}a_{32})X_2\\& + (a_{11}+a_{22}+a_{33})X_3 + a_{11}^2 u_1
\end{align*}
\end{example}

\section{Conclusion and Future work}
We have presented a new algorithm for finding globally identifiable reparametrizations of ODE models, which has wider applicability than the existing methods. 
 We have not systematically analyzed the scalability of our algorithm in relation to the number of parameters and state variables and degrees of nonlinearity. Just like with identifiability analysis~\cite{SIAN}, we do not see a clear dependence here besides non-polynomial rational models taking typically much longer computation time (e.g., NF$\kappa$B model is much bigger than Pharm, but the latter is rational and much less efficient for identifiability analysis). 

The two main bottlenecks for us are the computation of IO-equations and polynomial system solving, with the latter likely playing a bigger role. More and more efficient algorithms are being developed for both bottlenecks at the moment, see e.g.~\cite{structidjl,msolve}. 
The complexity of polynomial system solving is typically more affected by the number of variables. The number of variables in our approach is determined by the degrees of terms in Lie derivatives. The largest biological model we have seen that passes through the bottlenecks of computing IO-equations and polynomial system solving is the Akt pathway model~\cite{Fujita}, which is a non-linear model with 16~parameters, 9~state variables, 1~input, and 3~outputs, as in \cite[Example~11]{structidjl}.

Some biological models, such as the glucose-insulin model from~\cite{BIG2016}, involve high-degree but relatively sparse polynomials  in the right-hand side of the ODE system. Developing an approach that takes advantage of sparsity would significantly improve the efficiency of the current algorithm.  This model is the smallest one in size (besides having a term of degree $8$ in the ODE) we have seen that does not pass the bottleneck of polynomial system solving. The polynomial system we are solving has $150$ equations of average degree $5$ per equation, in $59$ unknowns, with rational function coefficients. Solving did not finish after $20$ days of sequential computing time, having consumed $120$ GB of RAM.

\section*{Acknowledgments} We thank Gleb Pogudin for detailed discussions of the main algorithm of the paper, CCiS at CUNY Queens College for the computational resources, and Julio Banga, Sebastian Falkensteiner, Gemma Massonis, Rafael Sendra and Alejandro Villaverde for useful suggestions. 
\bibliographystyle{IEEEtran}
\bibliography{bib.bib}
\end{document}